\documentclass[journal,twoside,web]{ieeecolor}
\usepackage{jsen}
\usepackage{cite}
\usepackage{amsmath,amssymb,amsfonts}
\usepackage{siunitx}
\usepackage{algorithmic}
\usepackage{graphicx}
\usepackage{wrapfig}
\usepackage{multirow}
\usepackage[english]{babel}
\usepackage[T1]{fontenc}
\usepackage[utf8x]{inputenc}
\usepackage[normalem]{ulem}
\usepackage[usestackEOL]{stackengine}
\DeclareSIUnit\permille{\text{\textperthousand}}
\usepackage[breaklinks,colorlinks=true,bookmarks=false,citecolor=blue,urlcolor=blue]{hyperref}
\def\BibTeX{{\rm B\kern-.05em{\sc i\kern-.025em b}\kern-.08em
    T\kern-.1667em\lower.7ex\hbox{E}\kern-.125emX}}

\definecolor{abstractbg}{rgb}{0.89804,0.94510,0.83137}
\begin{document}
\title{An Ultra-Sensitive Visible-IR Range Fiber Based Plasmonic Refractive Index Sensor }
\author{Mohammad Al Mahfuz, \IEEEmembership{Student Member, IEEE}, Sumaiya Afroj, Md. Azad Hossain, \IEEEmembership{Member, IEEE}, Md. Anwar Hossain, \IEEEmembership{Senior Member, IEEE},  Afiquer Rahman, and Md Selim Habib, \IEEEmembership{Senior Member, IEEE}
\thanks{Manuscript received January $\times\times$, 2024; $\times\times$, 2024; accepted $\times\times$, 2024. Date of publication $\times\times$,
2024; date of current version $\times\times$, 2024. 
(\emph{Corresponding author:
Md Selim Habib.})
}
\thanks{M. A. Mahfuz and M. S. Habib are with the Department of Electrical Engineering and Computer Science, Florida Institute of Technology, Melbourne, FL-32901, USA (e-mail: mmahfuz2024@my.fit.edu and mhabib@fit.edu). }
\thanks{S. Afroj is with the Department of Biomedical Engineering, Bangladesh University of Engineering and Technology, West Palashi 1000, Dhaka (e-mail: sumaiyaafrojmuna1997@gmail.com).}
\thanks{M. A. Azad and M. A. Mahfuz are with the Department of Electronics and Telecommunication Engineering, Chittagong University of Engineering and Technology, Raozan 4349, Chattogram (e-mail:  azad@cuet.ac.bd).}
\thanks{M. A. Hossain is with the Department of Electrical and Electronic Engineering, Bangladesh University of Business and Technology, Mirpur 1216, Dhaka 
(e-mail: dr.anwar@bubt.edu.bd).}
\thanks{A. Rahman is with the Department of Electronics and Telecommunication Engineering, Rajshahi University of Engineering and Technology, Kazla 6204, Rajshahi 
(e-mail: afiq.ruet18@gmail.com).}
}
\IEEEtitleabstractindextext{%
\fcolorbox{abstractbg}{abstractbg}{%
\begin{minipage}{\textwidth}%
\begin{wrapfigure}[12]{r}{3in}%
\includegraphics[width=3in]{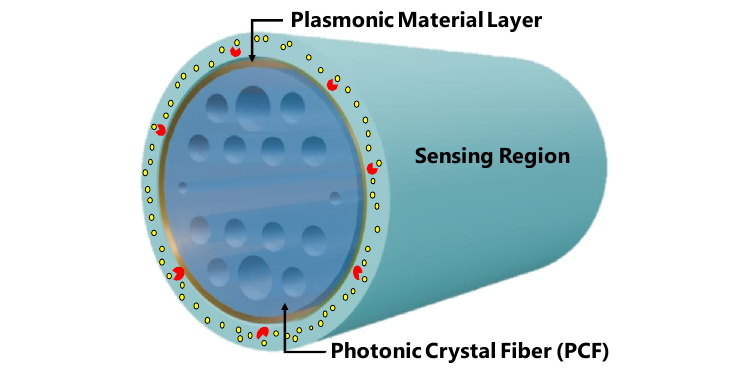}%
\end{wrapfigure}%

\begin{abstract}
Photonic crystal fiber (PCF)-based plasmonic sensors have gained considerable attention because of their highly sensitive performance and broad range of sensing regimes. In this work, a relatively simple ultra-sensitive PCF-based surface plasmon resonance (SPR) sensor has been proposed for detecting different analyte refractive indices (RIs) ranging from 1.33 to 1.43 over a wide range of wavelength spectrum spanning 0.55 \textmu m to 3.50 \textmu m. The comprehensive finite-element simulations indicate that it is possible to achieve remarkable sensing performances such as wavelength sensitivity (WS) and figure of merit (FOM) as high as 123,000 nm/RIU and 683 RIU$^{-1}$, respectively, and extremely low values of  wavelength resolution (WR) of 8.13 x 10$^{-8}$ RIU. In addition, a novel artificial neural network (ANN) model is proposed to be integrated into the practical setup in order to accurately predict the RIs by carefully examining the simulation data. The mean square error (MSE) and accuracy ($R^2$) values for the ANN model are found about 0.0097 and 0.9987, respectively, indicating the high prediction capability of the proposed ANN model. Due to its exceptional sensitivity and precise detection capabilities, the proposed device has the potential to serve as a viable option for sensing analyte refractive index (RI). Additionally, the sensor could be utilized for identifying cancerous cells and detecting urinary tract infections in humans.

\end{abstract}
\end{minipage}}}
\maketitle

\section{Introduction}
Breakthrough research on surface plasmon resonance (SPR) technology has been widely demonstrated in various lab-on-chip devices, sensors, filters, and so on. Utilizing SPR in optical sensing proves advantageous due to its effectiveness, real-time detection capabilities, and user-friendly operations \cite{das2023computational}, \cite{zhao2023research}. SPR technology holds promise for diverse applications such as food safety, security, medical testing, medical diagnostics, and biomolecular analyte detection \cite{ravindran2023recent, zhou2023biochemical}. Importantly, SPR sensors are appealing due to their reliability, efficiency, rapid response, effective light control capabilities, and label-free sensing, contributing to their widespread acceptance \cite{singh2023review}. However, traditional SPR sensors relying on slot waveguide \cite{laxmi2023nanophotonic}, fiber Bragg grating \cite{mishra2023fiber}, and prism coupling \cite{bolduc2009high}, tend to be costly and bulky. Overcoming these limitations, optical fiber-based SPR sensors emerge as compact and cost-effective alternatives \cite{jing2022performance}. The success rate of bio-analyte sensing surpasses that of conventional methods \cite{uniyal2023recent}, covering a range of applications including the detection of liquid analytes \cite{mumtaz2023numerical, daher2023novel}, gas detection \cite{liu2023copper}, and cancer detection \cite{kaur2022recent}. Plasmonic biosensors offering high-sensitivity refractive index detection have gained prominence. These encompass metal-based propagating eigenwaves RI sensors \cite{liu2015twin, liu2016multi, baiad2015concatenation}, nanoparticle-based localized surface plasmon resonance (LSPR) detectors \cite{li2015production, lodewijks2012boosting}, fano resonance RI sensors \cite{bao2015magnetic, lee2015ultrasensitive}, and hybrid plasmonic-photonic sensors \cite{baaske2014single, dantham2013label}. In contrast to conventional fiber optic sensors, in recent years there has been growing interest in combining SPR technology with photonic crystal fiber (PCF) for sensing applications owing to the design flexibility, improved sensitivity, portability, lightweight construction, remote sensing capabilities, high birefringence properties, and single-mode guidance \cite{zhao2019current, zhao2020applications, rifat2017photonic, al2019asymmetrical}. The miniaturization of the device is possible for PCF sensors due to its physical dimensions \cite{hasan2017spiral} and guiding properties can also be controlled by modifying the geometric parameters \cite{al2019highly}. In PCF-based plasmonic sensor, the coupling condition is greatly influenced by the surrounding environment, with the RI being a significant factor. Consequently, even slight alterations in the surrounding RI can potentially modify the resonance or coupling conditions. Therefore, detecting a shift in the resonance peak allows for the straightforward identification of an unknown analyte, as reported by Yang  \textit{et al.} \cite{yang2020anomalous}. 

In recent years, two types of PCF-based SPR sensing approaches have been studied namely internal sensing \cite{azman2024novel} and external sensing \cite{srivastava2022micro,haque2021highly,al2020dual,ramola2021design,sharif2023high}. In Ref. \cite{azman2024novel}, the authors studied an internal plasmonic sensor with dual channels, involving a selective coating of metal around the air-holes. This approach adds complexity to the device fabrication process, and the internal coating for both channels proves to be a time-consuming procedure. The performance of the sensor in Ref. \cite{azman2024novel} is comparatively low in which wavelength sensitivity (WS) and figure of merit (FOM) reached about 11,000 nm/RIU and 204 RIU$^{-1}$, respectively. In addition, the external sensing approach can potentially surpass the limitations of the internal sensing approach as the sensing medium and metal layer both are employed at the outer edge of the sensor and susceptible to the surrounding environment. A dual-core PCF external sensor was proposed by Mahfuz \textit{et al.} \cite{al2020dual}, which shows WS of 28,000 nm/RIU and FOM is about 2800 RIU$^{-1}$. Here, the FOM is high, however, the WS is comparatively low. Recently, Srivastava and colleagues \cite{srivastava2022micro} numerically reported a external micro-channel $D$-shaped PCF-SPR sensor that explores WS of 67,000 nm/RIU and FOM of 279.16 RIU$^{-1}$. The micro-channel sensor provides relatively low FOM. Another intricate $D$-shaped SPR-PCF sensor was proposed in Ref. \cite{haque2021highly} in which WS of 216,000 nm/RIU and FOM of 1200 RIU$^{-1}$ were obtained.  These are outstanding results, however, fabrication of $D$-shape structure is highly challenging and less practical since it requires precisely surface polishing. Besides, Sharif \textit{et al.} \cite{sharif2023high} proposed a circular PCF-SPR sensor for external sensing of samples which exhibits WS of 13,800 nm/RIU and FOM is about 470 RIU$^{-1}$. The sensing performance falls below the expected standard, and the range for sensing analyte refractive index is restricted, spanning from 1.29 to 1.34. The previous studies show either high WS and low FOM or vice-versa, along with fabrication challenges making them impractical for many real-world applications.

In our study, we propose a relatively simple an ultra-sensitive PCF-based SPR sensor considering external sensing technique with coating gold layer on the outer surface of the fiber which explores the large shifting of resonance wavelength for refractive index variation in the outer environment from the visible to the infrared regime. The performance of the sensor is carried out in terms of WS, WR, and FOM. In addition, a deep learning model is suggested using the dataset from simulation which can successfully predict the analyte refractive indices. The distinctive features of this study can be outlined as follows: (i) the fiber design is simple and cost-effective, (ii) practical implementation of the senor is feasible as it utilizes only two rings of air-holes, with metal coating applied to the external surface of the fiber, (iii) broad RI sensing range covering the detection regime from visible to the infrared spectrum spanning 0.55 \textmu m to 3.50 \textmu m, and (iv) remarkable sensing performances with better accuracy.

\section{Fiber Geometry and Theoretical Background}
\begin{figure}
  \begin{center}
  \includegraphics[width=2in]{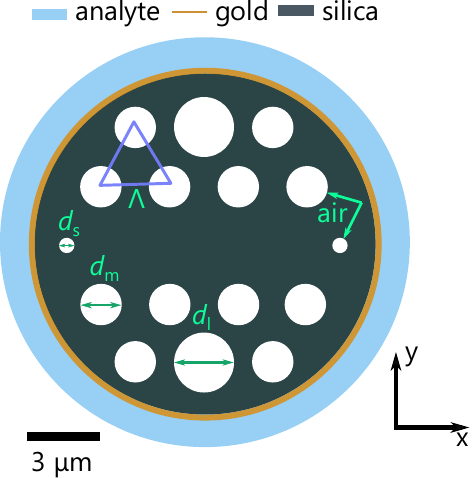}\\
   \caption{2D geometry of the proposed sensor. Geometrical parameters are considered as below, two adjacent air-holes center to center distance, $\varLambda$ = 2.8 µm, large air hole diameter, $d_{l}$ = 2.52 µm, medium air hole diameter, $d_{m}$ = 1.68 µm, small air hole diameter, $d_{s}$= 0.56 µm, gold layer thickness, $t_{g}$ = 35 nm, and analyte layer thickness, $t_{a}$ = 1.4 µm.}\label{fig:fig_1}
  \end{center}
\end{figure}

Fig. \ref{fig:fig_1} shows the two-dimensional view of the proposed PCF-based plasmonic RI sensor where the air-holes are arranged in two hexagonal rings on a silica ($\text{SiO}_\text{2}$) substrate. Two air-holes from the inner ring are omitted in the horizontal direction 
and two air-holes from the outer ring in the same direction are scaled down so that the incident electromagnetic light can precisely excite the surface electrons. In the orthogonal direction, two air-holes from the outer ring are scaled up which can assist to improve the performance of the sensor in the considered direction. In order to simplify the practical sensing approach, both the plasmonic material and sensing region are integrated onto the outer surface of the fiber, serving as an external sensing layer. The fiber material employed is fused silica, and the dielectric constants can be derived from the Sellmeier equation, as outlined in Ref.  \cite{tan1998determination}. Fused silica RI variation with the changes of temperature is 1.28x10$^{-5}/^o\text{c}$ only, therefore, the temperature effect is neglected in the simulations without extreme temperature fluctuations.

Plasmonic materials play a vital role in the PCF-SPR sensor, which is responsible for generating surface plasmon polariton (SPP) waves in the fiber-dielectric interface. Considering the characteristics of broad shift of resonance valley, inertness, and longer stability in the environment, gold (Au) is used as plasmonic material and the dielectric function of gold is taken from the Drude-Lorentz Model \cite{haider2018propagation}:

\begin{equation}
\epsilon_\text{Au}=\epsilon_{\infty}-\frac{\omega_{D}^{2}}{\omega(\omega+j\gamma_{D})}-\frac{\Delta\epsilon\Omega_{L}^{2}}{(\omega^{2}-\Omega_{L}^{2})+j\Gamma_{L}\omega},
\end{equation}\label{Eq:Eq_1} 

where $\varepsilon_{Au}$ = permittivity of Au and permittivity at high frequency $\varepsilon_{\text{\ensuremath{\infty}}}$ = 5.9673. Other constants are taken from Ref. \cite{haider2018propagation}. Nano layer of Au can be coated at the outer surface of the PCF by applying the atomic layer deposition (ALD) and chemical vapor deposition (CVD) method \cite{oviroh2019new}, \cite{gleason2015cvd}, \cite{al2020dual}. The sensing medium is placed at the external surface of the fiber and the sensor characteristic is changed with the changes of the outer environment. The target analyte can be detected with the functionalization of the Au film by varying the refractive index simultaneously.

\begin{figure}
  \begin{center}
  \includegraphics[width=3.5in]{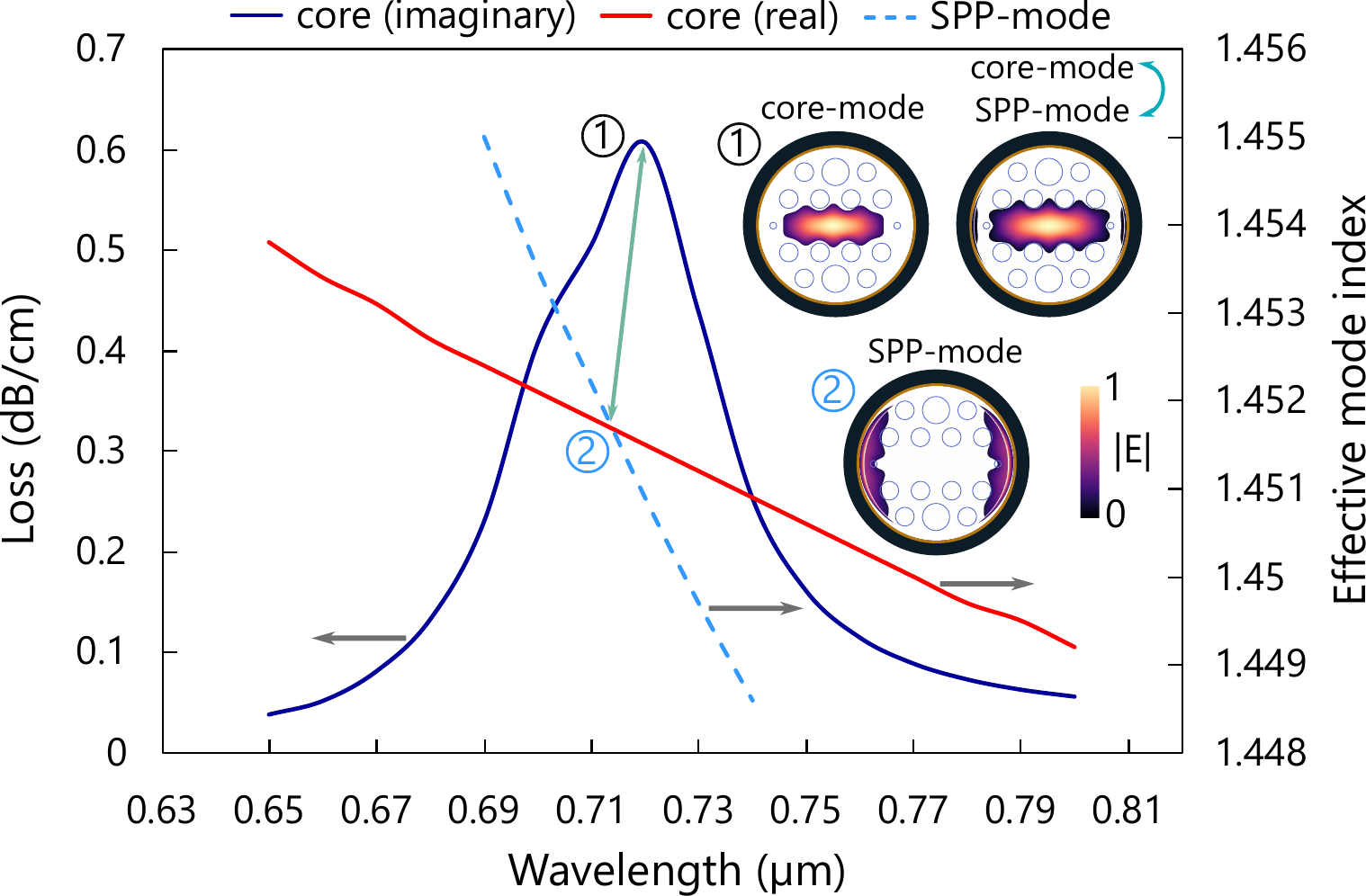}\\
   \caption{Phase matching properties for analyte RI of 1.39, where violet and red line show imaginary and real effective mode index for fundamental mode and broken light blue line indicates the SPP-mode, respectively. Inset figures represent the mode-field profiles for \textcircled{1} core-mode at wavelength 0.67 µm and coupling of core-mode and SPP-mode at 0.72 µm, and \textcircled{2} SPP-mode at 0.75 µm wavelength. The color bar shows the normalized mode-field profiles in a linear scale. The simulations are performed for $\varLambda$ = 2.8 µm, $d_{l}$ = 2.52 µm, $d_{m}$ = 1.68 µm, $d_{s}$= 0.56 µm, $t_{g}$ = 35 nm, and $t_{a}$ = 1.4 µm in $x$--polarization direction.}\label{fig:fig_2}
  \end{center}
\end{figure}

\section{Phase Matching Properties and EM Field Distributions}
PCF-based SPR sensor working principle is based on guided electromagnetic evanescent field with appropriately designed core-cladding geometry. Free electrons stimulate from the metallic surface upon arrival of propagated light guided through the fiber core region. A surface plasmon wave (SPW) is generated when the event of frequency matching occurs. These SPWs are strongly sensitive to the surrounding RIs. Local RIs can be detected by properly observing the resonance or spectral wavelength variations \cite{rifat2017photonic}. Fig. \ref{fig:fig_2} shows the phase-matching properties for local analyte RI of 1.39. It is evident from Fig. \ref{fig:fig_2} that for the core-guided mode the imaginary value of the effective mode index (violet) gradually increases, whereas the real value of the effective-mode index (red) gradually decreases with the increment of spectral wavelength. Besides, the real value of the effective-mode index (broken light blue) of SPP-mode gradually decreases and intersected at a wavelength of 0.72 µm. This is the phase matching point or resonance wavelength in which maximum energy transfers from the core-mode to the SPP-mode. The inset of Fig. \ref{fig:fig_2}:\textcircled{1} shows the mode field distributions of core-mode and the coupling between core-mode and SPP-mode, whereas the inset of Fig. \ref{fig:fig_2}:\textcircled{2} represents the SPP-mode for $x$--polarization.

\section{Simulation theory and sensor performance analysis}

The numerical simulations are performed through the finite-element method (FEM). To accurately model the fiber properties, a perfectly matched layer (PML) with optimized boundary conditions was incorporated outside the fiber domain, as outlined in \cite{habib2013proposal,habib2015low,habib2019single, cooper20232}. Extremely fine mesh sizes are considered during simulations. In order to get a better sensing response, the fiber parameters are optimized by tuning a particular parameter, while other parameters remain unchanged. Due to showing the improved sensing response, the parameters are selected as $\varLambda$ = 2.8 µm, $d_{l}$ = 2.52 µm, $d_{m}$ = 1.68 µm, $d_{s}$= 0.56 µm, $t_{g}$ = 35 nm, and $t_{a}$ = 1.4 µm. The proposed sensor performance is carried out using the confinement loss (CL) characteristics for various analytes in which the CL was calculated using the well-known equation as in Ref. \cite{hasan2014highly}.
The sensitivity of the proposed plasmonic sensor is observed using the wavelength interrogation method and the wavelength sensitivity (WS) of the proposed plasmonic RI sensor is calculated using the following equation \cite{hussain2023dual}:

\begin{equation}
S_{\lambda}=\frac{\varDelta\lambda_\text{peak}}{\Delta n_{a}} (\text{nm/RIU}),
\end{equation}\label{Eq:Eq_2}

where $\Delta\lambda_\text{peak}$ and $\Delta n_{a}$ indicate the wavelength shifting of the resonance peak and the RI variation of the two adjacent analytes, respectively. To observe the detection capability of the proposed sensor in the case of tiny analyte RIs variation, it is crucial to find out the sensor resolution which can be obtained from below equation \cite{al2020dual}:

\begin{equation}
\text{WR}=\frac{\Delta n_{a}\times\Delta\lambda_\text{min}}{\Delta\lambda_\text{peak}} (\text{RIU}),
\end{equation}\label{Eq:Eq_3}

where $\Delta n_{a}$, $\Delta\lambda_\text{peak}$, and $\Delta\lambda_\text{min}$ indicate the variation of two adjacent tiny RI variations, loss peak shifting and minimum resolution of the detector, respectively. During the measurement of sensor resolution external perturbation and instrumental noise are not considered \cite{al2020dual}. In determining the sensor's detection limit, it is equally crucial to compute the figure of merit (FOM), as expressed in the following equation \cite{feng2023design}:

\begin{equation}
\text{FOM}=\frac{\text{Wavelegth sensitivity}}{\text{FWHM}} (\text{RIU$^{-1}$}),
\end{equation}\label{Eq:Eq_4}

where FWHM stands for full width at half maximum. High FOM indicates the better detection limit of the sensor.

\begin{figure}
  \begin{center}
  \includegraphics[width=3.5in]{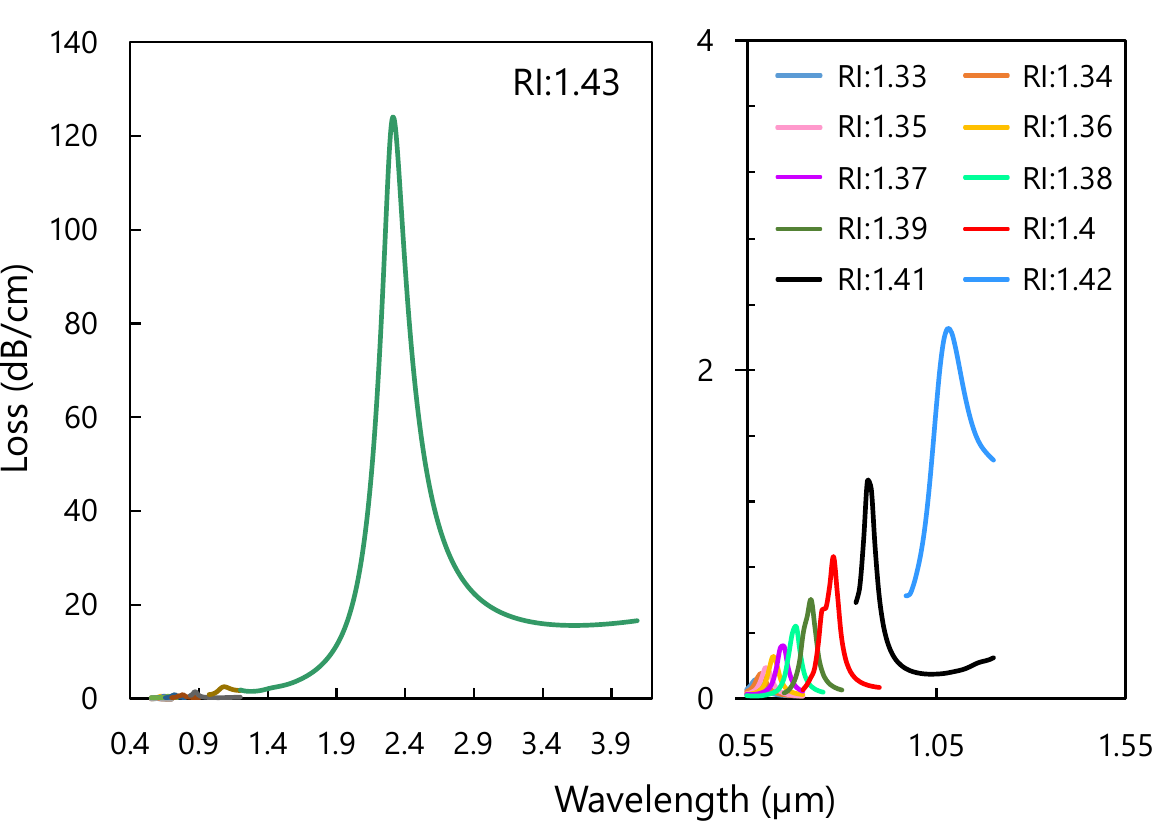}\\
   \caption{Confinement Loss curve for various analyte RIs from 1.33 to 1.43 in the wavelength between 0.5 to 4 µm (visible to IR range) using fiber geometrical parameters: $\varLambda$ = 2.8 µm, $d_{l}$ = 2.52 µm, $d_{m}$ = 1.68 µm, $d_{s}$= 0.56 µm, $t_{g}$ = 35 nm, and $t_{a}$ = 1.4 µm in $x$--polarization direction.}\label{fig:fig_3}
  \end{center}
\end{figure}

In this study, the observation of the sensor's performance focuses solely on the $x$--polarization mode due to the stronger coupling of the core SPP-mode. Fig. \ref{fig:fig_3} illustrates the CL spectra while varying the sample refractive indices within the range of 1.33 to 1.43. According to Fig. \ref{fig:fig_3}, CL curves growing and red-shifted simultaneously with the increasing of analyte RIs. This phenomenon occurred due to the RIs contrast reduction from core-mode to SPP-mode, hence, evanescent electromagnetic light coupling through the metal-dielectric interface efficiently increased and loss peak experienced an upward trend. It can be seen from Fig. \ref{fig:fig_3} that the lowest loss peak value of 0.11832 dB/cm appeared for RI of 1.33 at the wavelength of 0.58 µm and it reached the highest peak value of 124.12 dB/cm at wavelength of 2.31 µm for RI of 1.43. The other CL peak values are found 0.15378, 0.19371, 0.25697, 0.32117, 0.43864, 0.60741, 0.86943, 1.3286, and 2.2584 dB/cm, respectively, for analyte RIs from 1.34 to 1.42 in the wavelength range between 0.59 and 1.08 µm. Using Equation 2 the minimum wavelength sensitivity (WS) is obtained about 1000 nm/RIU for analyte RI of 1.33 and the maximum WS reached about 123,000 nm/RIU for sample RI of 1.42. Besides, the WR is measured about 8.13 x 10$^{-8}$ RIU, which indicates that the proposed sensor is capable of detecting very tiny changes of analyte RI even in the order of 10$^{-8}$ scale and maximum FOM obtained about 683 RIU$^{-1}$ for analyte RI of 1.42 using Eqs. (3) and (4), respectively. The performance of the proposed sensor is briefly tabulated in Table \ref{Table_1}.

\begin{table}[b!]
\centering
\caption{Details performance of the proposed RI sensor}
\label{Table_1}
\resizebox{\linewidth}{!}{%
\begin{tabular}{ccccccc}
\hline \hline
\textbf{\begin{tabular}[c]{@{}c@{}}Analyte\\  RI\end{tabular}} & \textbf{\begin{tabular}[c]{@{}c@{}}Loss peak \\ (dB/cm)\end{tabular}} & \textbf{\begin{tabular}[c]{@{}c@{}}$\text{W}_\text{peak}$\\  (µm)\end{tabular}} & \textbf{\begin{tabular}[c]{@{}c@{}}$\text{S}_\text{peak}$\\  (nm)\end{tabular}} & \textbf{\begin{tabular}[c]{@{}c@{}}WS\\ (nm/RIU)\end{tabular}} & \textbf{\begin{tabular}[c]{@{}c@{}}FWHM\\ (nm)\end{tabular}} & \textbf{\begin{tabular}[c]{@{}c@{}}FOM\\ (1/RIU)\end{tabular}} \\ \hline
1.33 & 0.11832 & 0.58 & 10 & 1000 & 50 & 20 \\ \hline
1.34 & 0.15378 & 0.59 & 10 & 1000 & 40 & 25 \\ \hline
1.35 & 0.19371 & 0.6 & 20 & 2000 & 40 & 50 \\ \hline
1.36 & 0.25697 & 0.62 & 30 & 3000 & 40 & 75 \\ \hline
1.37 & 0.32117 & 0.65 & 30 & 3000 & 40 & 75 \\ \hline
1.38 & 0.43864 & 0.68 & 40 & 4000 & 40 & 100 \\ \hline
1.39 & 0.60741 & 0.72 & 60 & 6000 & 40 & 150 \\ \hline
1.4 & 0.86943 & 0.78 & 90 & 9000 & 50 & 180 \\ \hline
1.41 & 1.3286 & 0.87 & 210 & 21000 & 50 & 420 \\ \hline
1.42 & 2.2584 & 1.08 & 1230 & 123000 & 180 & 683 \\ \hline
1.43 & 124.12 & 2.31 & -- & -- & -- & -- \\ \hline
\end{tabular}%
}
\end{table}

\begin{figure}
  \begin{center}
  \includegraphics[width=3.5in]{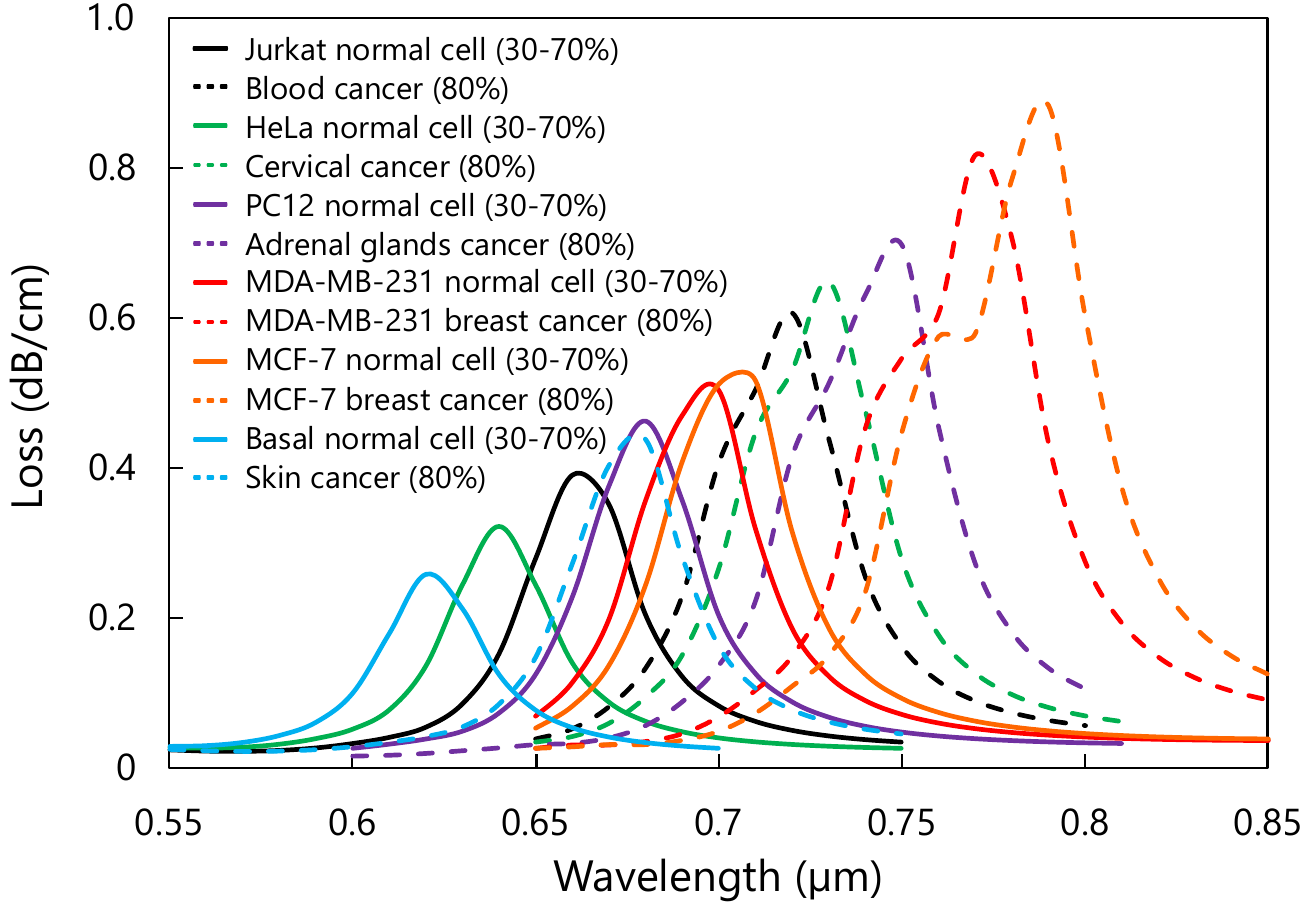}\\
   \caption{Resonance peak shifting for seven different biological cells and the dotted curves indicate the cancer-affected cell whereas the smooth curves represent normal cells.}\label{fig:fig_4}
  \end{center}
\end{figure} 

In this study, the detectable RI range is 1.33-1.43 which span is significant for various cell biology research and disease diagnoses \cite{liu2016cell}.
Fig. \ref{fig:fig_4} illustrates that the proposed sensor can potentially detect the various cancer-affected cells by observing the resonance valley shift from the normal cell. The CL peaks appeared at 0.62, 0.64, 0.66, 0.68, 0.7, and 0.71 µm wavelength for normal cells of basal (30-70\%), HeLa (30-70\%), Jurkat (30-70\%), PC12 (30-70\%), MDA-MB-231 (30-70\%), and MCF 7 (30-70\%), respectively. However, when the cells are affected by skin (80\%), cervical (80\%), blood (80\%), adrenal glands (80\%), and breast (80\%) cancers then the zenith peaks are moved to the longer wavelength which are found at 0.68, 0.73, 0.72, 0.75, 0.78, and 0.79 µm with showing WS of 6000, 7000, 6000, 7000, 8000, and 8000 nm/RIU, respectively. Importantly, the sensor exhibits low loss characteristics (below 1 dB/cm), making it an excellent choice for detecting various cancerous cells in the visible regime. Here, the relevant RIs values are taken from the following reference \cite{jabin2019surface}.

\begin{figure}
  \begin{center}
  \includegraphics[width=3.5in]{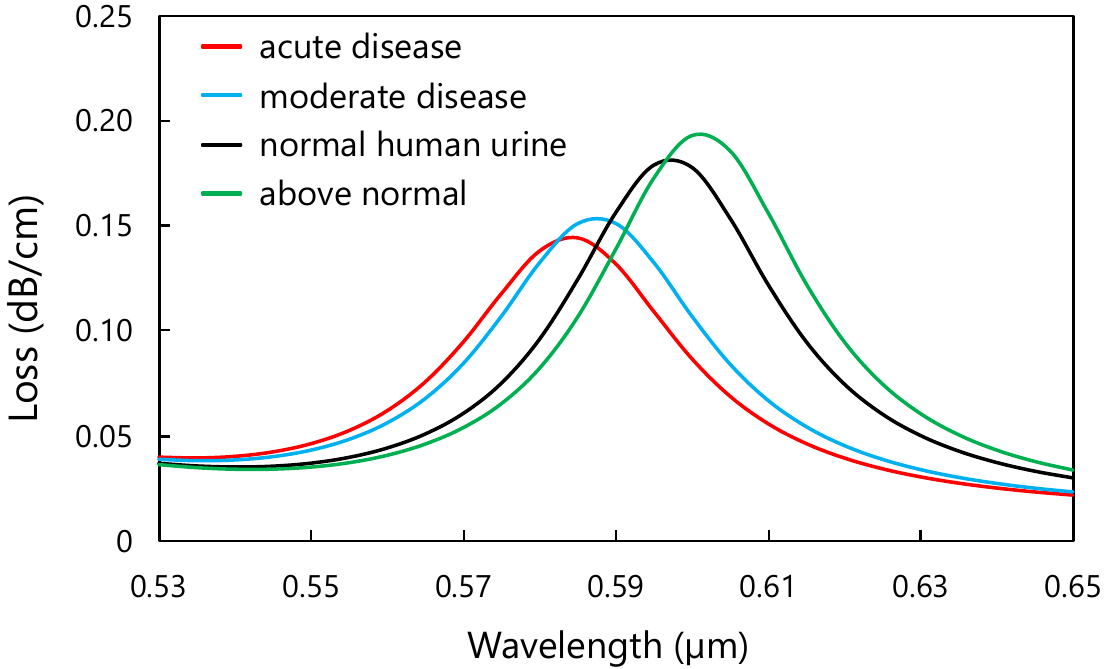}\\
   \caption{Observation of resonance peak shifting due to human urine infections. The RIs of normal human urine, moderate disease, acute disease, and above normal are 1.3464, 1.3396, 1.337, and 1.3489, respectively.}\label{fig:fig_5}
  \end{center}
\end{figure}

\begin{table}[]
\centering
\caption{Performance comparison with the reported PCF based SPR sensor}
\label{Table_2}
\resizebox{\linewidth}{!}{%
\begin{tabular}{cccccc}
\hline \hline
\textbf{Refs.} & \textbf{\begin{tabular}[c]{@{}c@{}}Sensing \\ type\end{tabular}} & \textbf{\begin{tabular}[c]{@{}c@{}}WS\\ (nm/RIU)\end{tabular}} & \textbf{\begin{tabular}[c]{@{}c@{}}WR\\ (RIU)\end{tabular}} & \textbf{\begin{tabular}[c]{@{}c@{}}FOM\\ (1/RIU)\end{tabular}}  & \textbf{\begin{tabular}[c]{@{}c@{}}Detection \\ range\end{tabular}} \\ \hline
\cite{hasan2017spiral} & External & 4600 & 2.17×10$^{-5}$ & -- & 1.33--1.38 \\ \hline
\cite{azman2024novel} & Internal & 11000 & 9.09×10$^{-5}$ & 204 & 1.33--1.41 \\ \hline
\cite{haque2021highly} & External & 216,000 & 4.63×10$^{-7}$ & 1200 & 1.33--1.42 \\ \hline
\cite{sharif2023high} & External & 13800 & 7.24×10$^{-6}$ & 470 & 1.29--1.34 \\ \hline
\cite{haider2018propagation} & External & 30000 & 3.33×10$^{-6}$ & 508 & 1.33--1.39 \\ \hline
\cite{hussain2023dual} & External & 24300 & 4.12×10$^{-6}$ & 120 & 1.10--1.45 \\ \hline
\cite{rahman2021highly} & External & 45800 & -- & 635 & 1.36--1.42 \\ \hline
\cite{haque2018surface} & Microchannel & 20000 & 5×10$^{-6}$ & -- & 1.18--1.36 \\ \hline
\cite{rifat2018highly} & Internal & 11000 & 9.1×10$^{-6}$ & 407 & 1.33--1.42 \\ \hline
Proposed & External & 123,000 & 8.13×10$^{-8}$ & 683 & 1.33--1.43 \\ \hline
\end{tabular}%
}
\end{table}

For the further applications of the sensor, in this work, we have figured out the sensing response in terms of human urine disease detection as depicted in Fig. \ref{fig:fig_5}. From Fig. \ref{fig:fig_5}, it is noticeable that the resonance valley is changed from its normal position when it is affected by disease. For normal urine (RI = 1.3464) the CL  peak is found at 0.595 µm wavelength and after being affected by disease it is shifted to the shorter wavelength at 0.585 and 0.59 µm for acute and moderate disease showing WS about 1000 and 500 nm/RIU, respectively. In the case of above normal, it is shifted to the longer wavelength of 0.6 µm, hence, the WS is 500 nm/RIU. In this specific detection study, the loss peaks remain below 0.2 dB/cm which is comparatively lower than the \cite{manickam2023numerical}, hence it can be a good candidate for biochemical sensing applications. In this specific detection study, the RIs values are taken from the following reference \cite{wolf1966aqueous}.
To validate the numerical results of the proposed sensor the performance is benchmarked with the published literature and tabulated in Table \ref{Table_2}. The proposed sensor can be practically implemented by using the stack-and-draw fabrication technique \cite{rahman2021highly}.

\section{ANN model for prediction of RIs from data obtained using the sensor}

This section introduces an artificial neural network (ANN) model integrated at the termination of the spectrum analyzer within the practical configuration of the sensor. The primary objective is to enhance the precision of the bio-analyte RI prediction technique subsequent to the acquisition of resonance wavelength and peak confinement loss. The ANN structure is displayed in Fig. \ref{fig:fig_6}, configured to take resonance wavelength and confinement loss as inputs and yield the corresponding RI as output. The distinct advantage of employing this model lies in its ability to significantly reduce human intervention in RI detection. The dataset for model training encompasses a spectrum of RI values ranging from 1.33 to 1.43, strategically chosen to reflect diverse bio-samples. For each RI, data points consist of resonance wavelength, four neighboring wavelengths, and corresponding confinement losses, ensuring resilience against fabrication errors leading to deviations in resonance wavelength. The input features and the output target are extracted, and in MaxScaler\cite{misra2019impact} is applied to normalize both input and output variables. The dataset is partitioned into a 7:3 ratio for training and validation purposes. The neural network model is constructed using the Sequential API from Keras \cite{ketkar2017introduction}. It consists of an input layer with 2 nodes corresponding to the number of features, followed by three dense layers with 512, 32, and 160 nodes, respectively, and varying activation functions (swish\cite{gustineli2022survey} and relu \cite{agarap2018deep}). Each hidden layer in a neural network uses activation functions to conduct a weighted sum or nonlinear processing on the values received from the prior layer. The output layer receives the updated value from the last concealed layer. Stochastic Gradient Descent (SGD), as outlined by Xu et al. \cite{xu2021interfacing}, is critical in maximizing all parameters via the connections made in the feed-forward process:
\begin{figure}
  \begin{center}
  \includegraphics[width=3.5in]{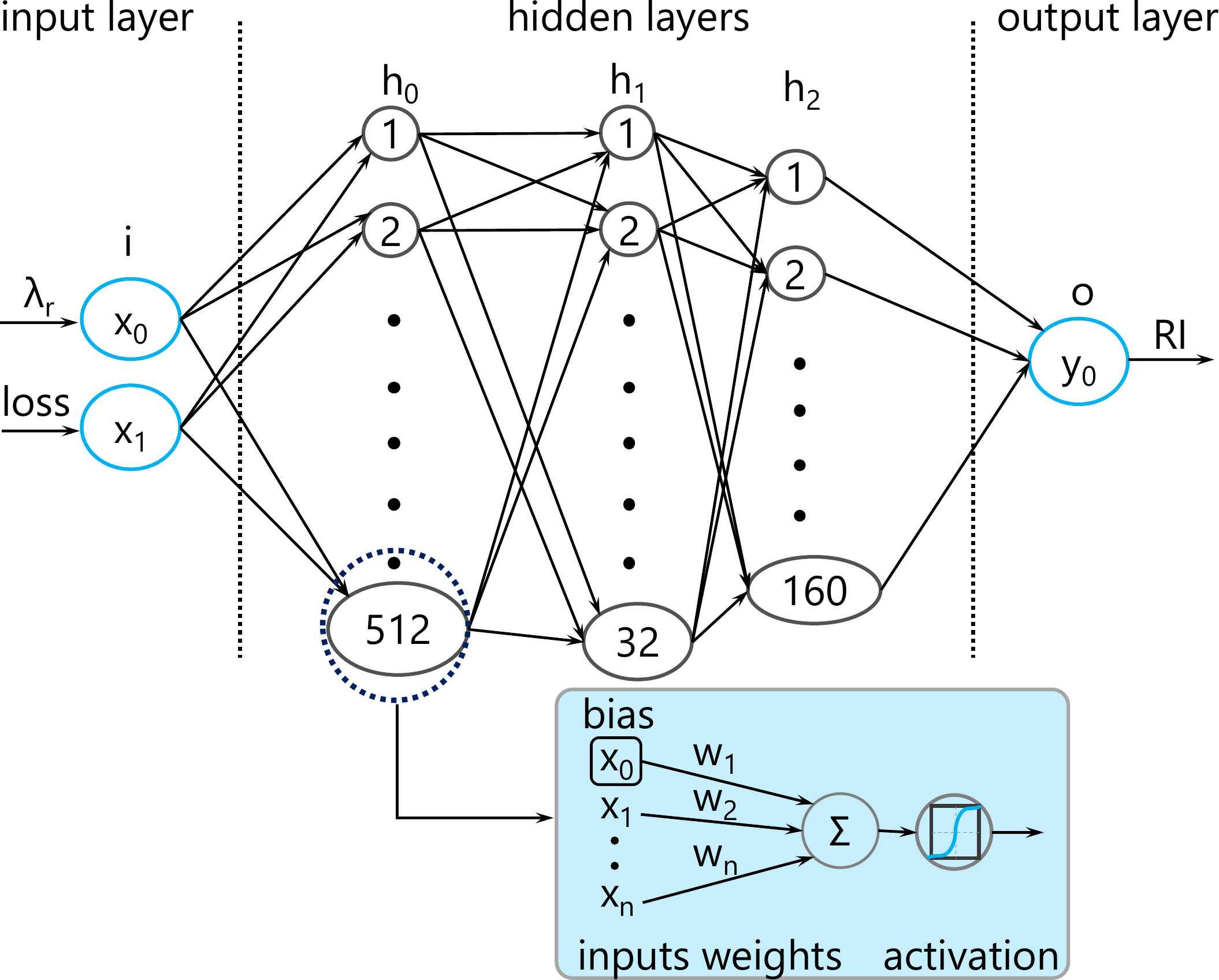}\\
   \caption{Structure of the neural network consisting of 3 hidden layers with different numbers of neurons in each layer. The input layer consists of two nodes: resonance wavelength ($\lambda_\text{r}$) and confinement loss. The output layer has one node: refractive index (RI). The mechanism of the activation function at each node is depicted in the smaller block.}\label{fig:fig_6}
  \end{center}
\end{figure}

\begin{equation}
  \text{weight}' = \text{weight} - \text{lr} \cdot \frac{\partial \text{loss}(Y, \hat{Y})}{\partial \text{weight}}.  
\end{equation}

\begin{equation}
    \text{bias}' = \text{bias} - \text{lr} \cdot \frac{\partial \text{loss}(Y, \hat{Y})}{\partial \text{bias}}.
\end{equation}

\begin{figure}
  \begin{center}
  \includegraphics[width=3.2in]{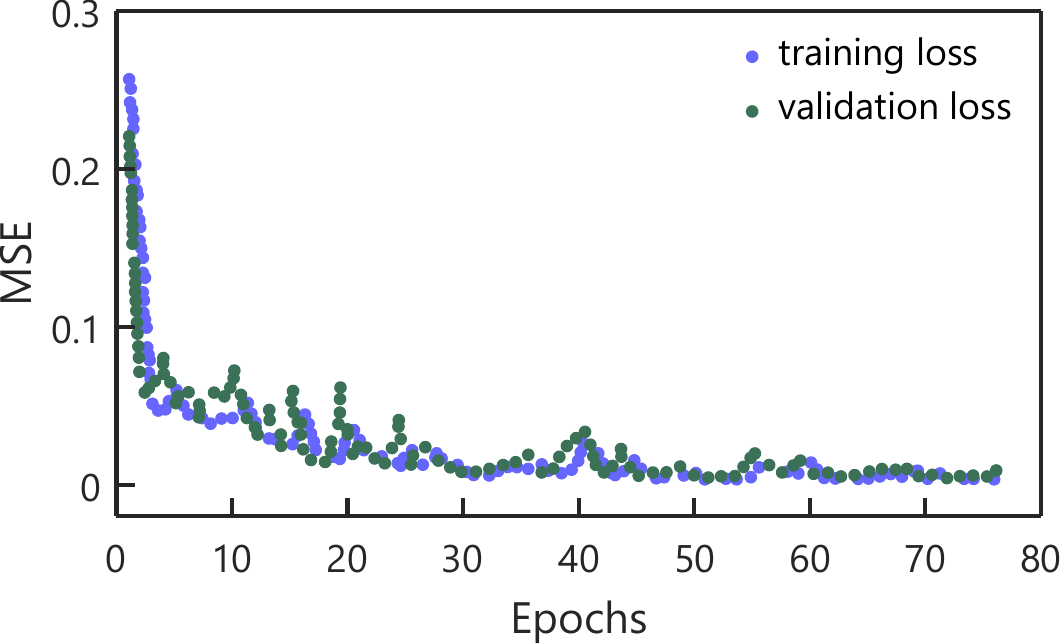}\\
   \caption{MSE during the training process against each epoch. The loss function stabilizes reaching the value of 0.0097 at around 80 epochs. The violet dot represents the training loss and the validation loss is depicted by green dot.}\label{fig:fig_7}
  \end{center}
\end{figure}
In this context, the learning rate (lr) is an important component. Neurons undergo nonlinear transformations via the activation function, as described by Leshno et al.\cite{leshno1993multilayer}. Because of this property, ANNs can approximate any desired function. In our method, we chose a very complex neural network rather than the traditional ANN topologies commonly used for regression problems. The complexity was added to improve prediction accuracy. 

\begin{figure}
  \begin{center}
  \includegraphics[width=3.5in]{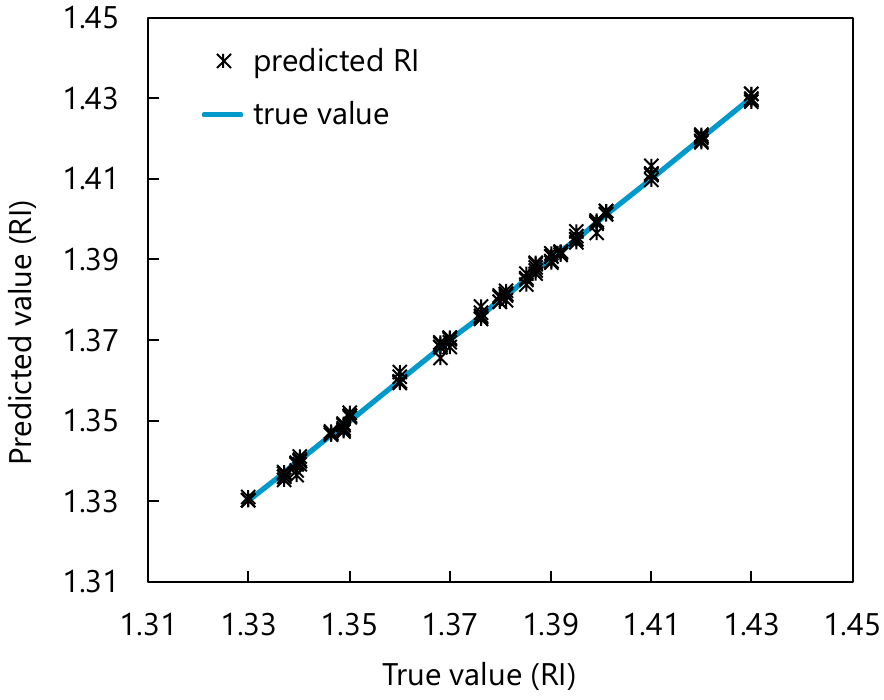}\\
   \caption{Actual RI vs predicted RI demonstrating the predictability of the model. Predicted RIs are marked with black markers and the blue line represents the actual RI. The predicted RIs do not deviate much from the actual RI for inputs consisting of wavelength and magnitude of confinement loss in the training range.}\label{fig:fig_8}
  \end{center}
\end{figure}

Notably, the last layer is made up of a single node with a linear activation function, making it an excellent choice for regression-oriented applications. All the hyper parameters for the ANN is tuned on the basis of trial and error method. The model is compiled using the Nadam \cite{wang2017chinese} optimizer with a specified learning rate and Mean squared error (MSE) as the loss function. As mentioned in Ref. \cite{bunke1984estimators}, the goal of this adjustment procedure is to reduce the MSE between the expected and actual output values. Formulating a loss function specifically for the regression problem is as follows:

\begin{equation}
   \text{MSE} =\frac{1}{n}\sum_{i=0}^{n}(y_i -\hat{y}_i)^{2}, 
\end{equation}

where $n$ is the number of data points and $y_i$ and $\hat{y}$ is the actual and predicted result respectively. Early stopping is implemented with a patience of 100 epochs to prevent over fitting. Training and validation loss over epochs, providing a visual representation of the model's learning progress is evident in Fig. \ref{fig:fig_7}. The MSE for the test dataset is obtained to be 0.0097. Fig. \ref{fig:fig_8} presents the predicted RIs in comparison to the actual values, underscoring the regression model's accuracy. The $R^2$ value for the predicted result obtained is 0.9987.

Some of the prediction results and their corresponding absolute percentage error is tabulated in Table \ref{Table_3}. The testing points used were completely new to the neural network and were not part of the training datasets. Because the network is tested on data that it has not encountered during the learning phase, this method ensures an unbiased evaluation of its prediction skills.

\begin{table}[t!]
\centering
\caption{Some input, output predicted from neural network and  their corresponding absolute percentage error}
\label{Table_3}
\resizebox{\linewidth}{!}{%
\begin{tabular}{ccccc}
\hline \hline
\multicolumn{2}{c}{\textbf{Neural network input}} & \multirow{2}{*}{\textbf{\begin{tabular}[c]{@{}c@{}}Predicted \\ RI\end{tabular}}} & \multirow{2}{*}{\textbf{\begin{tabular}[c]{@{}c@{}}Actual \\ RI\end{tabular}}} & \multirow{2}{*}{\textbf{$\Bigl|\text{Error (\%)}\Bigr|$}} \\ \cline{1-2}
\textbf{\begin{tabular}[c]{@{}c@{}}RW\\  (µm)\end{tabular}} & \textbf{\begin{tabular}[c]{@{}c@{}}CL\\(dB/cm)\end{tabular}} &  &  &  \\ \hline
0.580 & 0.11832 & 1.330000 & 1.3300 & 0 \\ \hline
0.585 & 0.14449 & 1.337010 & \begin{tabular}[c]{@{}c@{}}1.3370\\  (Acute disease)\end{tabular} & 7.48$\cdot$$10^{-4}$ \\ \hline
0.595 & 0.17909 & 1.346900 & \begin{tabular}[c]{@{}c@{}}1.3464\\  (Normal urine)\end{tabular} & 3.70$\cdot$$10^{-2}$ \\ \hline
0.660 & 0.38897 & 1.375954 & \begin{tabular}[c]{@{}c@{}}1.3760\\  (Normal cell jurkat)\end{tabular} & 3.34$\cdot$$10^{-3}$ \\ \hline
0.750 & 0.69604 & 1.394826 & \begin{tabular}[c]{@{}c@{}}1.3950 \\ (Adrenal glands cancer)\end{tabular} & 1.24$\cdot$$10^{-2}$ \\ \hline
0.780 & 0.70629 & 1.399046 & \begin{tabular}[c]{@{}c@{}}1.3990\\  (MDA-MB-231)\end{tabular} & 3.28$\cdot$$10^{-3}$ \\ \hline
2.310 & 124.120 & 1.430759 & 1.4300 & 5.30$\cdot$$10^{-2}$ \\ \hline
\end{tabular}%
}
\end{table}

\section{Conclusion}
In summary, a relatively simple and highly sensitive PCF-SPR sensor is proposed and studied using finite-element modeling. The overall performance of the sensor is remarkable as it exhibits maximum wavelength sensitivity of 123,000 nm/RIU, which potentially cover wide wavelength spectrum from the visible to infrared range that span from 0.55 \textmu m to 3.50 \textmu m. It also offered maximum wavelength resolution of 8.13$\times$10$^{-8}$ RIU which indicate the proposed sensor can easily detect the sample analyte RI variation even in 10$^{-8}$ scale. Furthermore, the sensor exhibits an improved detection capability, as indicated by the FOM value of approximately 683 RIU$^{-1}$. The RIs range which can be detectable by this sensor are from 1.33 to 1.43, that includes various cancerous cells, biomolecules, biochemicals, and so on. Finally, a noble ANN model is designed that can predict analyte RIs with high accuracy and low loss. Due to the promising sensing response with comparatively lower loss function, the proposed PCF sensor can be significantly applicable for bio-analyte detection in a lab-on-a-chip platform.


\bibliography{References}

\bibliographystyle{ieeetr}
\vfill
\end{document}